\documentclass[10pt,conference]{IEEEtran}
\IEEEoverridecommandlockouts
\usepackage{enumitem}
\usepackage{url}

\usepackage{commath, color}
\usepackage{amssymb}
\usepackage{comment}
\usepackage[ruled,vlined]{algorithm2e}
\usepackage{cite}
\usepackage{amsmath,amssymb,amsfonts}
\usepackage{algorithmic}
\usepackage{graphicx}
\usepackage{textcomp}
\usepackage{xcolor}
\usepackage{comment}
\usepackage{subfigure}
\usepackage[utf8]{inputenc}

\usepackage{siunitx}
\def\BibTeX{{\rm B\kern-.05em{\sc i\kern-.025em b}\kern-.08em
    T\kern-.1667em\lower.7ex\hbox{E}\kern-.125emX}}
\DeclareUnicodeCharacter{FB01}{fi}
\begin{document}
\title{Improving Spatio-Temporal Understanding of Particulate Matter using Low-Cost IoT Sensors}
\author{
\IEEEauthorblockN{C. Rajashekar Reddy, T. Mukku, A. Dwivedi, A. Rout, \\ S. Chaudhari, K. Vemuri, K. S. Rajan, A. M. Hussain
\IEEEauthorblockA{
International Institute of Information Technology-Hyderabad (IIIT-H), India\\ Emails: \{rajashekar.reddy, ayush.dwivedi\}@research.iiit.ac.in, tanmai.mukku@students.iiit.ac.in\\  \{sachin.c, kvemuri, rajan, aftab.hussain\}@iiit.ac.in}}
}

\maketitle

\begin{abstract}
Current air pollution monitoring systems are bulky and expensive resulting in a very sparse deployment. In addition, the data from these monitoring stations may not be easily accessible. This paper focuses on studying the dense deployment based air pollution monitoring using IoT enabled low-cost sensor nodes. For this, total nine low-cost IoT nodes monitoring particulate matter (PM), which is one of the most dominant pollutants, are deployed in a small educational campus in Indian city of Hyderabad. Out of these, eight IoT nodes were developed at IIIT-H while one was bought off the shelf.
A web based dashboard website is developed to easily monitor the real-time PM values. The data is collected from these nodes for more than five months. Different analyses such as correlation and spatial interpolation are done on the data to understand efficacy of dense deployment in better understanding the spatial variability and time-dependent changes to the local pollution indicators.
\end{abstract}


\section{Introduction}
Air pollution is one of the world’s largest environmental causes of diseases and premature death \cite{Lancet2018}. Out of different air pollutants, particulate matter (PM) has been identified as one of the most dangerous pollutants. Because of long-term exposure of PM, every year millions of people die and many more become seriously ill with cardiovascular and respiratory diseases \cite{Tagle2020}. The issues are more aggravated in a developing country like India, where large sections of the population are exposed to high levels of PM levels \cite{Pant2019}. With increasing urbanization, the situation is only going to get worse. Recent study in \cite{Wu2020} has also shown that a small increase in long-term exposure to PM2.5 leads to a large increase in COVID-19 death rate. Therefore, it is important to develop tools for monitoring PM so that timely decisions can be made. In this paper, the focus is particularly on monitoring mass concentrations of PM2.5 (\emph{fine PM} or particles with aerodynamic diameter less than 2.5 \si{\micro \meter}) and PM10 (\emph{coarse PM} or particles with aerodynamic diameter between 2.5 \si{\micro \meter} and 10 \si{\micro \meter}) as these two PMs are mostly linked with human health impacts \cite{Pant2019}.

Traditionally, PM monitoring is done using scientific-grade devices such as beta attenuation monitor (BAM) and tapered element oscillating microbalance (TEOM) deployed by pollution controlling boards and other governmental agencies. Although these systems are reliable and accurate, there are two important issues. First is that these systems are expensive, large and bulky, which leads to sparse deployment. For example, there are six monitoring stations deployed by Central Pollution Control Board (CPCB) in the Indian city of Hyderabad, which is spread over an area of 650 \si{\kilo \meter \squared} \cite{CPCB}. Also, these stations provide temporally more coarse data (hourly or daily). This in turn leads to low spatio-temporal resolution which is not enough to understand the exposure of citizens to pollution, which is non-uniformly distributed over the city. Second issue is that the measured pollution data at the monitoring stations and estimates at other locations are not readily available \cite{Brienza2015}. This lack of access to information results in lack of awareness among the citizens regarding the pollution in their area of residence or frequently visited locations such as home, office, schools and gardens. 

Low-cost portable sensors along with internet of things (IoT) can overcome the above two issues of traditional monitoring systems. The low-cost portable ambient sensors provide a huge opportunity in increasing the spatio-temporal resolution of the air pollution information and are even able to verify, fine-tune or improve the existing ambient air quality models \cite{Badura2018}. It has been shown in \cite{Brienza2015} that a low-cost monitoring system, which is not as accurate as a traditional and expensive one, can still provide reliable indications about air quality in a local area. IoT along with dense deployment of such low-cost sensors can provide real-time access of pollution data with high spatio-temporal resolution. Government and citizens can use this information to identify pollution hot-spots so that timely and localized decisions can be made regarding reducing and preventing air pollution.

There has been some work on PM monitoring in the literature \cite{Badura2018, Johnston2019, Pant2019, Tagle2020}. In\cite{Badura2018, Johnston2019}, the performances of different low-cost optical PM2.5 sensors such as Nova SDS011, Winsen ZH03A, Plantower PMS7003, Honeywell HPMA115S0 and Alphasense OPC-N2 have been evaluated. Authors in \cite{Pant2019} presented regulatory PM2.5 and PM10 data availability along with the current status of the national monitoring networks and plans. In \cite{Tagle2020} and \cite{Johnston2019}, very few (six and three, respectively) IoT nodes measuring PM2.5 and PM10 were deployed in different geographical regions of Santiago, Chile, and Southampton, UK respectively, to examine the suitability of low-cost sensors for PM monitoring in urban environment. However, there is a dearth of actual deployment and measurements of dense IoT network to map fine spatio-temporal PM variations, which is precisely the focus of this paper.

This paper focuses on studying the dense deployment based air pollution monitoring using IoT enabled low-cost sensor nodes in Indian urban conditions. For this, eight sensor nodes measuring PM2.5 and PM10 are developed and deployed in IIIT-H campus, which is 0.267 \si{\kilo \meter \squared}. A web-based dashboard is developed to easily monitor the real-time air pollution\footnote{The website is live but the historic data, schematics and codes will be made public once the paper is published.}. One of the eight deployed nodes is co-located with commercially available and factory-calibrated sensor node with a view of calibrating developed sensor nodes.  The data is collected from these nine nodes for approximately five months. Correlation analysis is done to understand correlation between different nodes in this denser (than traditional) deployment. For spatial interpolation, inverse distance weighing (IDW) scheme is used on these nodes for the data collected before and during the bursting of firecrackers on the main night of Diwali (one of the most popular festivals in India) to show the variability pattern in a small campus, hot spot detection and need for a dense deployment to provide better local pollution indicators. 

The paper is organized as follows. In Section \ref{sec:Methodology}, details on IoT network development and deployment along with measurement campaign are presented followed by data analysis tools in \ref{sec:DataProcessing}. Section  \ref{sec:Results} present the results while Section \ref{sec:Conclusion} concludes the paper.

\section{IoT Network Implementation and Field Measurements} \label{sec:Methodology}
\subsection{Sensor Node Implementation}
\begin{figure}[bhtp]
    \centering
    \subfigure[Block architecture]{
    {\includegraphics[width=0.9\columnwidth]{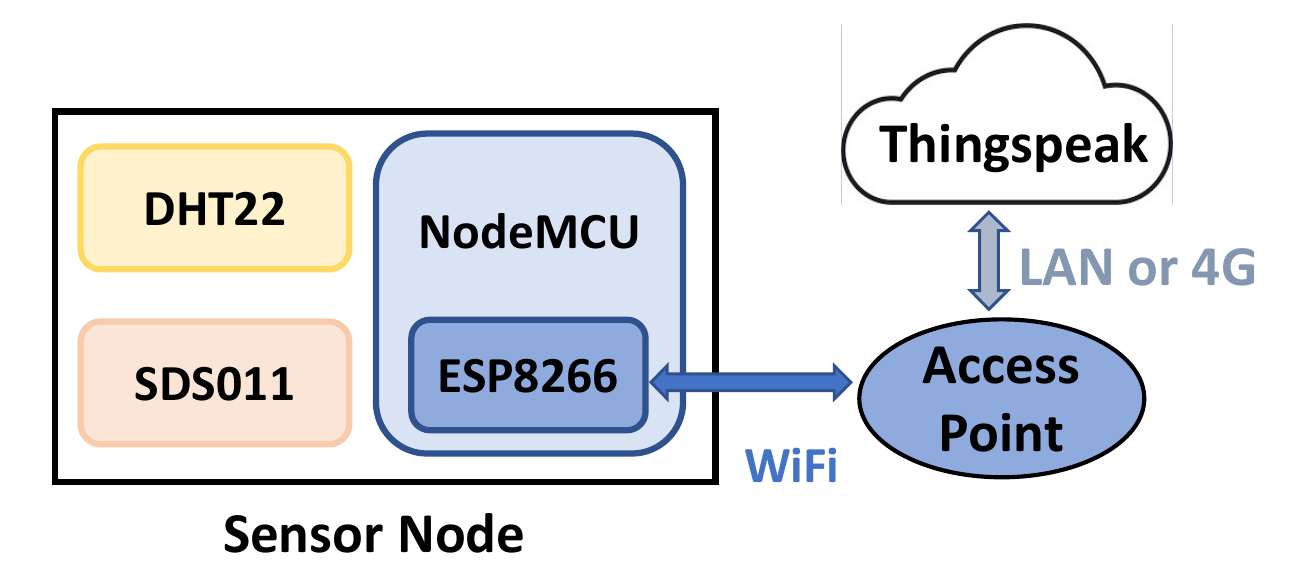}}
    \label{BlockDiagram}
    }
    \\
    \subfigure[Circuit diagram]{
    {\includegraphics[width=0.7\columnwidth]{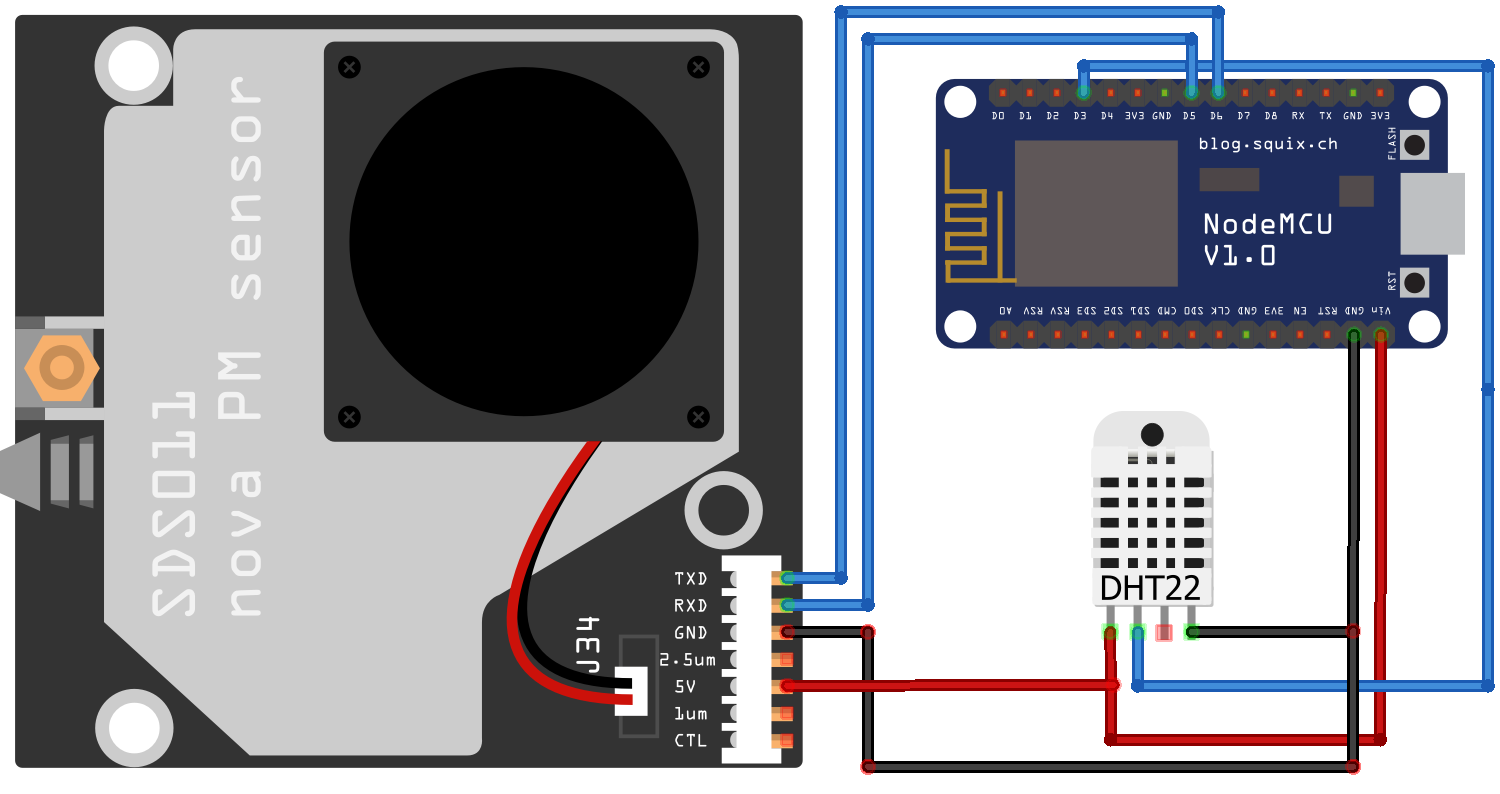}}
    \label{Circuit}
    }
    \caption{Block and circuit diagrams of sensor node developed at IIIT-H for monitoring PM values.}
    \label{BlockAndCktDiagram}
\end{figure}

Figs. \ref{BlockDiagram} and \ref{Circuit} show the block architecture and circuit diagram, respectively, of the PM monitoring sensor node developed at IIIT-H. Each node consists of ESP8266 based NodeMCU microcontroller and sensors for PM, temperature and humidity. The specifications of the sensors used are given in Table \ref{SensorSpecifications}. Nova PM SDS011 which is light scattering principle based sensor, has been used for PM2.5 and PM10 measurements as it has been shown to have best performance among several low cost PM sensors in terms of closeness to the expensive and accurate beta attenuation mass (BAM) and reproducibility among different SDS011 units \cite{Badura2018}. Since the light scattering based PM sensors do not perform reliably at extreme temperature and humidity conditions, DHT22 is used to monitor these parameters for reliability of SDS011 sensor readings. 

NodeMCU samples data from the sensors and transmits it periodically via WiFi to \emph{ThingSpeak} \cite{ThingSpeak}, which is a cloud based IoT platform for storing and processing data using MATLAB, 
for logging the data. NodeMCU uses on-chip ESP8266 module to connect to available WiFi access point for internet connection. NodeMCU samples the Nova PM SDS011 sensor for PM2.5 and PM10 in \si{\micro\gram\per\cubic\meter} and DHT22 sensor for environmental conditions temperature and relative humidity in \si{\celsius} and \% respectively at a sampling rate of 15 seconds and the network delay added for the communication with the server. The connections are made using a PCB printed and designed at IIIT-H for stability of the connectors between the sensors and the microcontroller.

\begin{table}[bthp]
\caption{Specifications of sensors used in the developed sensor node.}
\centering
\begin{tabular} {|p{2.1cm}|p{1.7cm}|p{1.4cm}|p{1.9cm}|}
 \hline
 Sensor & Parameter & Resolution & Relative error \\ [0.5ex] 
 \hline
 SDS011 \cite{SDS011} & PM2.5, PM10 & 0.3\si{\micro\gram\per\cubic\meter} & Max. of ±15\%, ±10 \si{\micro\gram\per\cubic\meter} \\ 
\hline
 DHT22 \cite{DHT22} & Temperature & 0.1\si{\celsius} & ±0.5\si{\celsius} \\
 \hline
 DHT22 \cite{DHT22}& Humidity & 0.1\% & ±2\% \\ [0.5ex]  
 \hline
\end{tabular}
\label{SensorSpecifications}
\end{table}

\begin{figure}[htbp]
\centerline{\includegraphics[width=0.6\columnwidth]{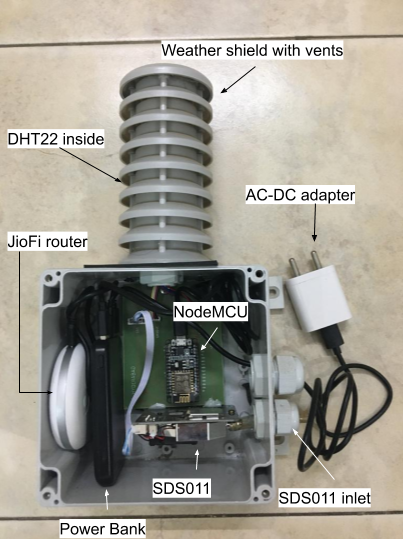}}
\caption{Outdoor air pollution node.}
\label{Node}
\end{figure}

Fig. \ref{Node} shows a deployment ready sensor node which consists of sensors, a NodeMCU, a 5000 \si{\milli\ampere\hour} power bank, 4G based portable WiFi routers (VoLTE-based JioFi JMR1040 \cite{JioFi}) and a weather shield. Power bank is needed for power backup in case of any fluctuations or drop in the power supply. A weather shield design with vents shown is used along to cater the ambient air flow requirements of  DHT22 for temperature and humidity. The components are enclosed in a poly carbonate box of IP65 rating as the deployment is outdoors. IP65 enclosures offer complete protection against dust particles and a good level of protection against water. 4G based WiFi router shown in the figure is not common to all the nodes deployed and is used only when the node is deployed outside campus WiFi coverage. 

\subsection{IoT Network Deployment}
\begin{figure}[tbp]
\centerline{\includegraphics[width=0.5\textwidth]{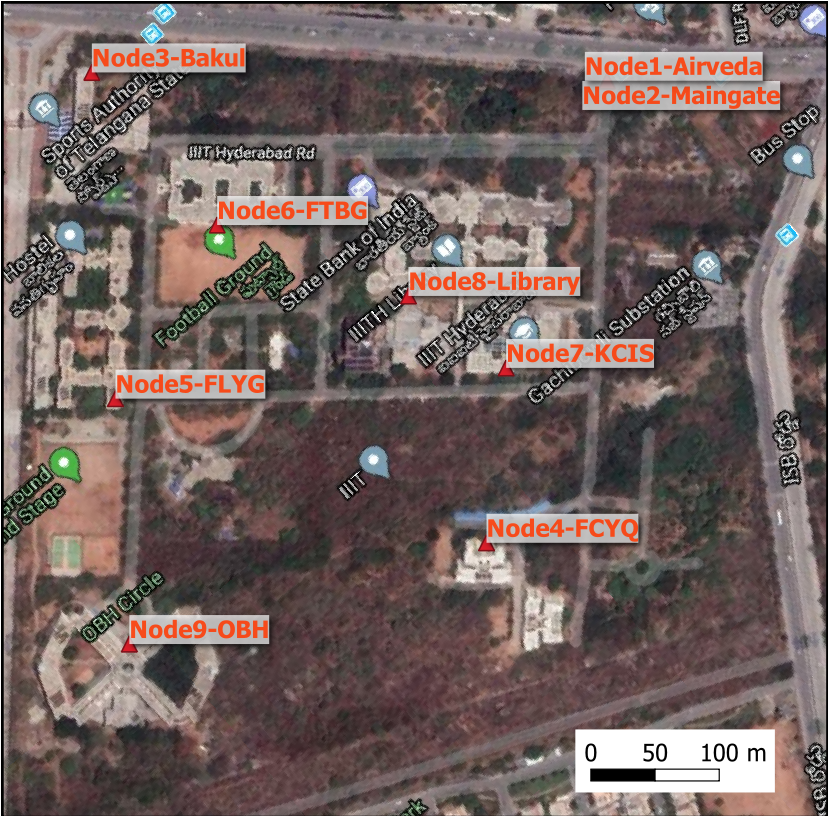}}
\caption{Sensor deployment in IIIT-H campus}
\label{Deployment}
\end{figure}

The prototype deployment and measurement region is the IIIT-H campus, Gachibowli, Hyderabad, India as shown in Fig.\ref{Deployment}. The area of the measurement region is 66 acres (0.267 \si{\kilo \meter \squared}). In this small campus, eight nodes developed at IIIT-H were deployed outdoors at locations shown in Fig. \ref{Deployment}. The figure also shows the notations and numbering of the nodes, which will be followed for the rest of the paper. Before deploying, these eight nodes were collocated in a lab and measured data for seven days to ensure that none of the devices is too deviant from the bunch. The deployment period of the nodes has been from 26 October 2019 to 10 April 2020 (more than 5 months). 

In addition to the eight nodes, a ninth node was also deployed by buying off-the-shelf commercial node from Airveda \cite{Airveda}. This node was factory calibrated with respect to BAM and has been used as a reference node for our nodes in this work. This node is denoted as Node1-Airveda and is collocated with Node2-MainGate as shown in Fig. \ref{Deployment}.

All the nodes are connected to continuous power supply. Nodes 4,6 and 8 are connected to the WiFi provided by the access points which are part of the campus WiFi network. Node 7 could connect to the campus WiFi network but with weak signal strength, which sometimes resulted in connection outages and data loss. To avoid this and strengthen the WiFi signal, a NodeMCU has been deployed in appropriate location as a WiFi repeater. Nodes 2,3,5 and 9 are out of the campus WiFi coverage and have been equipped with individual 4G based portable JioFi WiFi routers for internet connectivity. Node1-Airveda is using WiFi provided by JioFi connected to Node2-MainGate since these two nodes are collocated.

Each of the eight IIIT nodes (i.e., nodes 2 to 9) uploads the sampled sensor data, namely PM2.5, PM10, temperature and relative humidity to individual channels created on the ThingSpeak server using \textsc{GET} method of the HTTP protocol. Node1-Airveda uses ESP8266 for WiFi communication and uploads data to Airveda server. The same data is retrieved using Airveda application program interface (API) and saved in a separate channel in the ThingSpeak server.

\subsection{Development of web-based dashboard}
\begin{figure}[bthp]
\centerline{\includegraphics[width=0.4\textwidth]{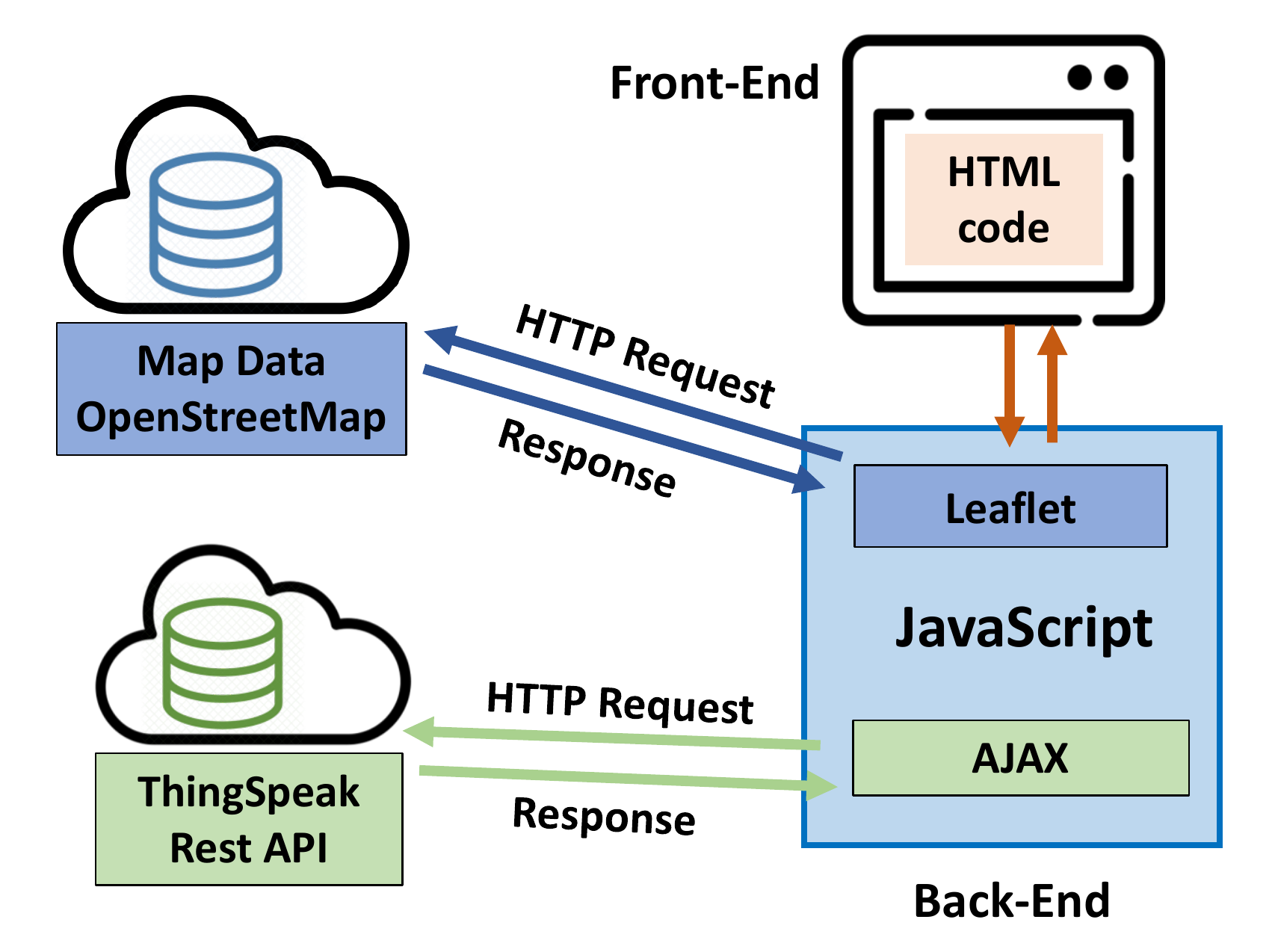}}
\caption{Process flow of website development for real-time PM monitoring.}
\label{Website}
\end{figure}


The website developed for displaying the real-time PM values is hosted at the address  \url{https:/spcrc.iiit.ac.in/air/}. Fig. \ref{Website} shows the process flow of the web-bas. The webpage is designed such that the data is fetched from the ThingSpeak server and is displayed on the webpage on an open source map \emph{OpenStreetMap} \cite{OpenStreetMap}. The front-end of the webpage is designed in \emph{hyper text markup language (HTML)} and back-end is designed in \emph{Javascript}. To get the map data, we used Javascript library \emph{Leaflet} \cite{Leaflet}. To get data from ThingSpeak, another Javascript library \emph{asynchronous Javascript and XML (AJAX)} is used, which allows us to get data from the ThingSpeak API using the GET function.  After we get the data from the ThingSpeak API, the data is averaged and a colour is associated with the data value. Next, using Leaflet functions, the marker colour and information on the map is set. The process then goes to sleep. The complete process is repeated every 5 minutes. Note that this dashboard does not show Node1-Airveda at the moment as it can be viewed on the Airveda webpage or app by adding the station ID.

\section{{Data Processing Methods}}\label{sec:DataProcessing}
\subsection{Data Cleaning and Preprocessing}
The following tasks were done to convert the raw data received from the sensor nodes into a usable data set:
\begin{itemize}
    \item It is essential to remove the outliers in a raw dataset as there are few extreme values that deviate from other samples in the data, which might be a result of several factors. Data cleaning can be done using clustering based outlier detection, which is a well known unsupervised method used extensively. In this paper, density based clustering algorithm in \cite{Celik2011} has been employed to identify the outliers and the vectors with outlier have been dropped. Environmental conditions such as temperature and humidity can affect the working of laser based PM sensors like SDS011. For example, there is overestimation of PM values at higher humidity. As such, these points also act as outliers and the corresponding vectors are removed using the density based clustering.
    \item Data averaging helps to look past random fluctuation and see the central trend of a data set. The sensor used in the PM measurements has a relative error of 15\% so  averaging the data helps to smooth the time series curve.
\end{itemize}

\subsection{Analysis tools}
\subsubsection{Quantile-Quantile plots}
The quantile-quantile plot or QQ plot is an analysis tool to assess if a pair of data variables' population possibly came from same distribution or not. A QQ plot is a scatterplot created by plotting two sets of quantiles against one another. If both sets of quantiles have come from the same distribution, the scatter plot form a line that’s roughly straight. Many distributional aspects like shifts in location, scale, symmetry, and the presence of outliers can be detected from these plots. For two data sets that come from similar populations whose data distribution functions differ only by shifts mentioned earlier, the data points lie along a straight line displaced either up or down from the 45-degree reference line. QQ plots help us understand the distributional features of the data sets and provide necessary confidence for assumptions for further analysis.

\subsubsection{Correlation Analysis}
Correlation is a bivariate analysis that measures the strength of association between two variables and the direction of the relationship. The correlation coefficient is a statistic tool used to measure the extent of the relationship between variables when compared in pairs. In terms of the strength of the relationship, the value of the correlation coefficient varies between +1 and -1. There are several types of correlation coefficients such as Pearson and Kendall. Pearson’s correlation is one of the most commonly used correlation coefficient but makes several assumptions on the data such as normally distributed variables, linearly related variables, complete absence of outliers and homoscedasticity. On the other hand, Kendall's tau doesn't require the above mentioned assumptions and is more suitable for the work in this paper. Kendall's tau ($\tau$), which is a non-parametric rank-based measure of dependence is defined as
\begin{equation}
    \tau = \frac{n_c-n_d}{n_c+n_d},\nonumber
\end{equation}
where $n_c$ and $n_d$ are the numbers of concordant pairs and discordant pairs respectively. For a given pair $(x_i, y_i)$ and $(x_j, y_j)$, let us define $z = (x_i - x_j)(y_i - y_j)$. This pair is concordant if $z > 0$ and discordant if $z < 0$. 


\subsubsection{Spatial Interpolation}
It is not practical to deploy and measure PM values at every location in the area of interest. However, using nearest measurement point to approximate the PM value at a location of interest may lead to erroneous results given the variability of pollution levels and weather in different locations in an urban environment. This can be mitigated by using \emph{spatial interpolation} to estimate the PM values at unmeasured locations using known values at the measurement locations. In this paper, we have used IDW, which is one of the simplest and popular deterministic spatial interpolation technique \cite{Burrough1998}. IDW follows the principle that the nodes that are closer to the location of estimation will have more impact than the ones which are farther away. IDW uses linearly weighted combination of the measured values at the nodes to estimate the parameter at the location of interest. The weight corresponding to a node is a function of inverse distance between the location of the node and the location of the estimate. In this paper, weights have been chosen to be inverse distance squared.


\section{Analysis and Results} \label{sec:Results}
The following analyses were applied on the obtained data set after cleaning and preprocessing: QQ plots, time series plots, correlation analysis and spatial analysis.

\subsection{Quantile-quantile plots (QQplots)}
\begin{figure*}[htbp]
    \centering
    \subfigure[PM 2.5]
    {\includegraphics[width=0.9\columnwidth]{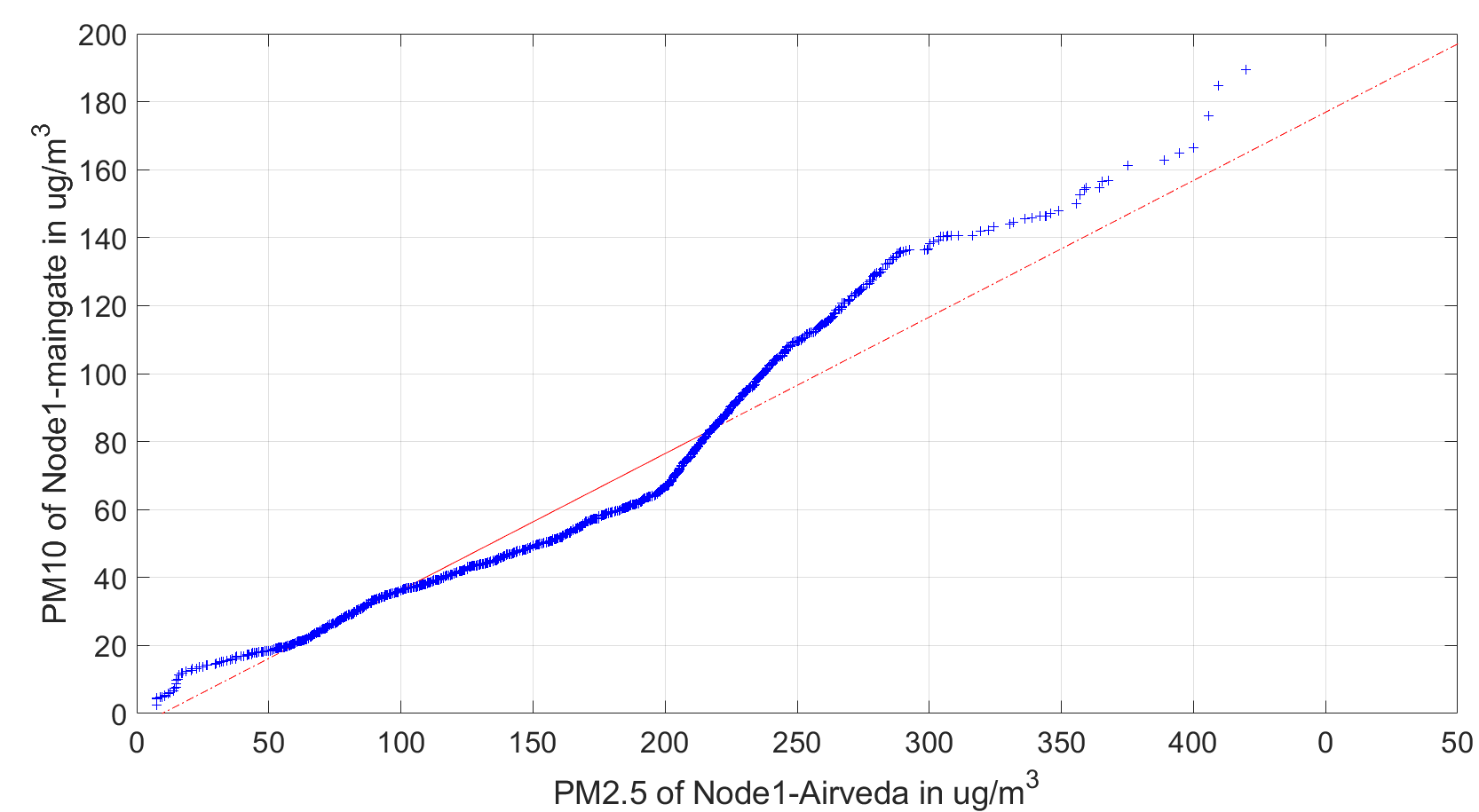}
    \label{QQ-PM25}
    }
    \quad
    \subfigure[PM 10]
    {\includegraphics[width=0.9\columnwidth]{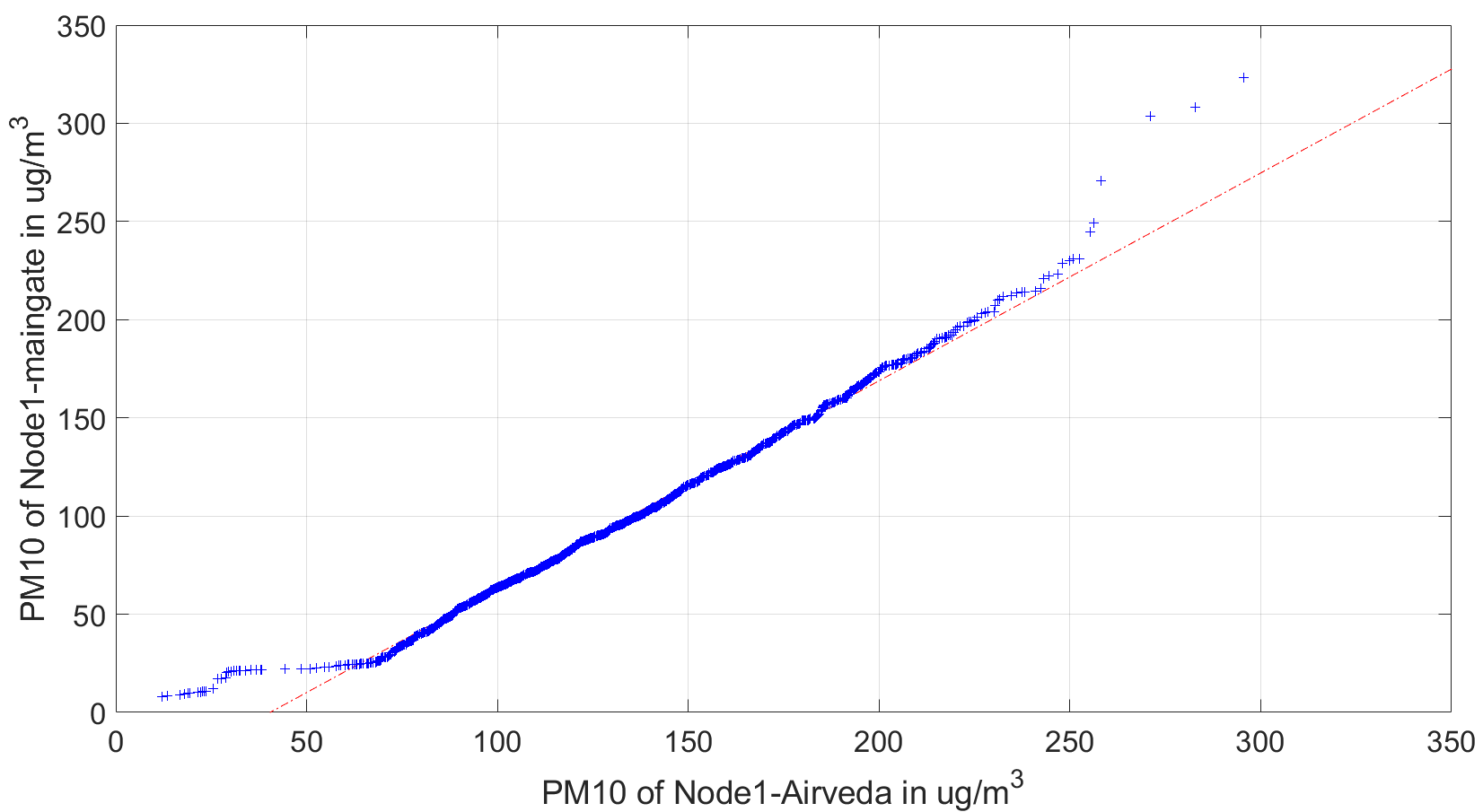}
    \label{QQ-PM10}
    }
    \caption{QQ plots for PM2.5 and PM10 between co-located nodes at the main gate, i.e., Node1-Airveda and Node2-Maingate.}
    \label{QQPlots}
\end{figure*}

QQ plots have been used on the two co-located nodes Node1-Airveda and Node2-Maingate to verify the distribution similarity. Node1-Airveda is an air quality monitoring device from Airveda which has been tested against the standard PM sensor BAM monitor and Node2-Maingate is the sensor node developed at IIIT-H. QQplots have been plotted with one-hour averaged data for PM2.5  in Fig. \ref{QQ-PM25} and for PM10 in Fig. \ref{QQ-PM10} with Node1-Airveda on horizontal axis and  Node2-Maingate on the vertical axis. The plots show linearity for most part with most of the sample points close to straight line with high density and very few points deviating from the linear relationship for both PM2.5 and PM10 samples. In the case of PM2.5 few deviating points belong to the higher end of the distribution while in the case of PM10 samples, few deviations can be seen at both lower and higher ends of the distribution. From the plots, it is safe to assume that the populations of the data samples of Node1-Airveda and Node2-Maingate follow a similar distribution with very few samples deviating.

\subsection{Time series plots for PM2.5 and PM10}
\begin{figure*}[htbp]
    \centering
    \subfigure[PM2.5]{
    {\includegraphics[width=0.7\textwidth]{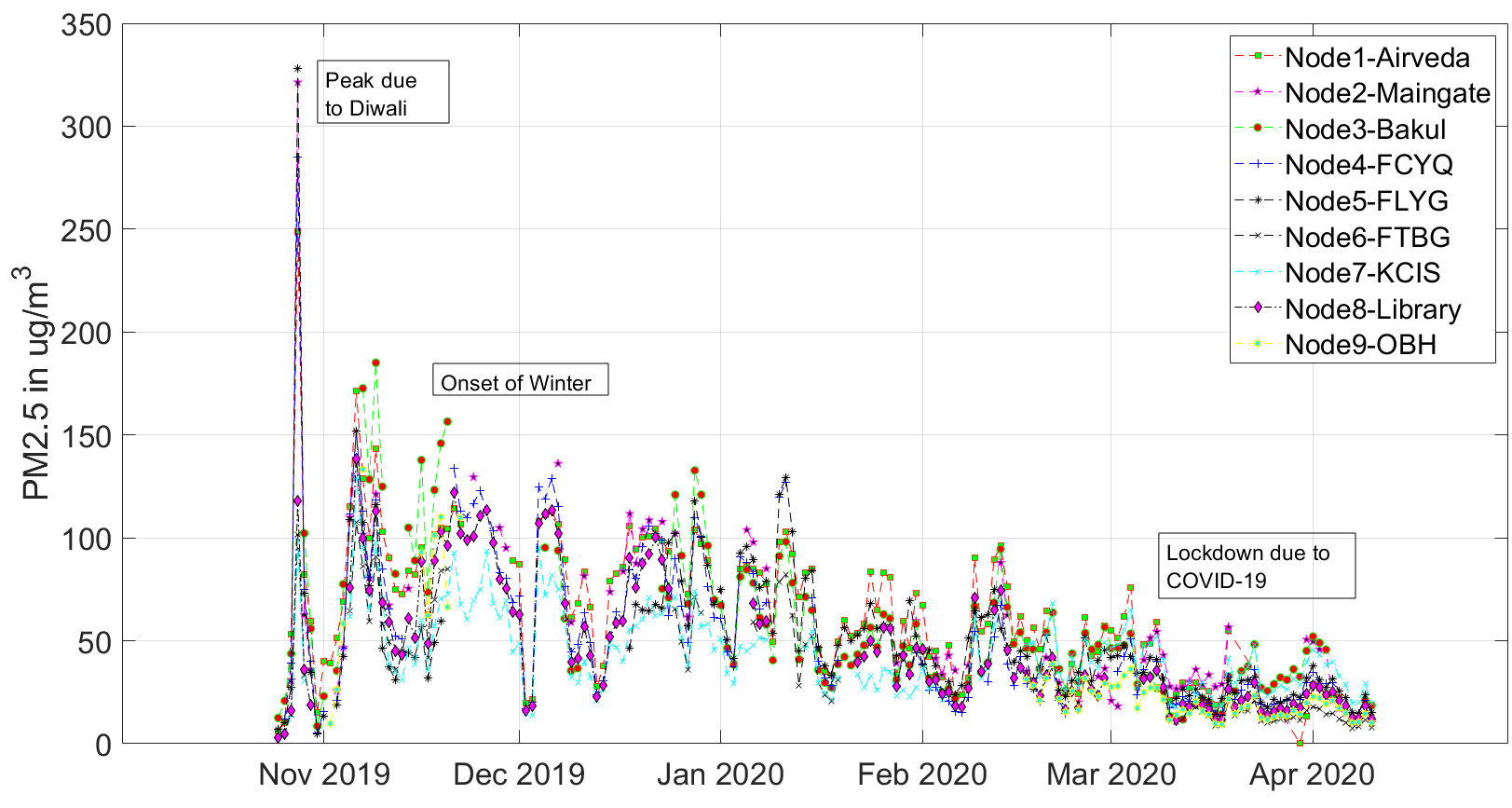}}
    \label{LinePlotPM25}
    }
    \quad
    \subfigure[PM10]{
    {\includegraphics[width=0.7\textwidth]{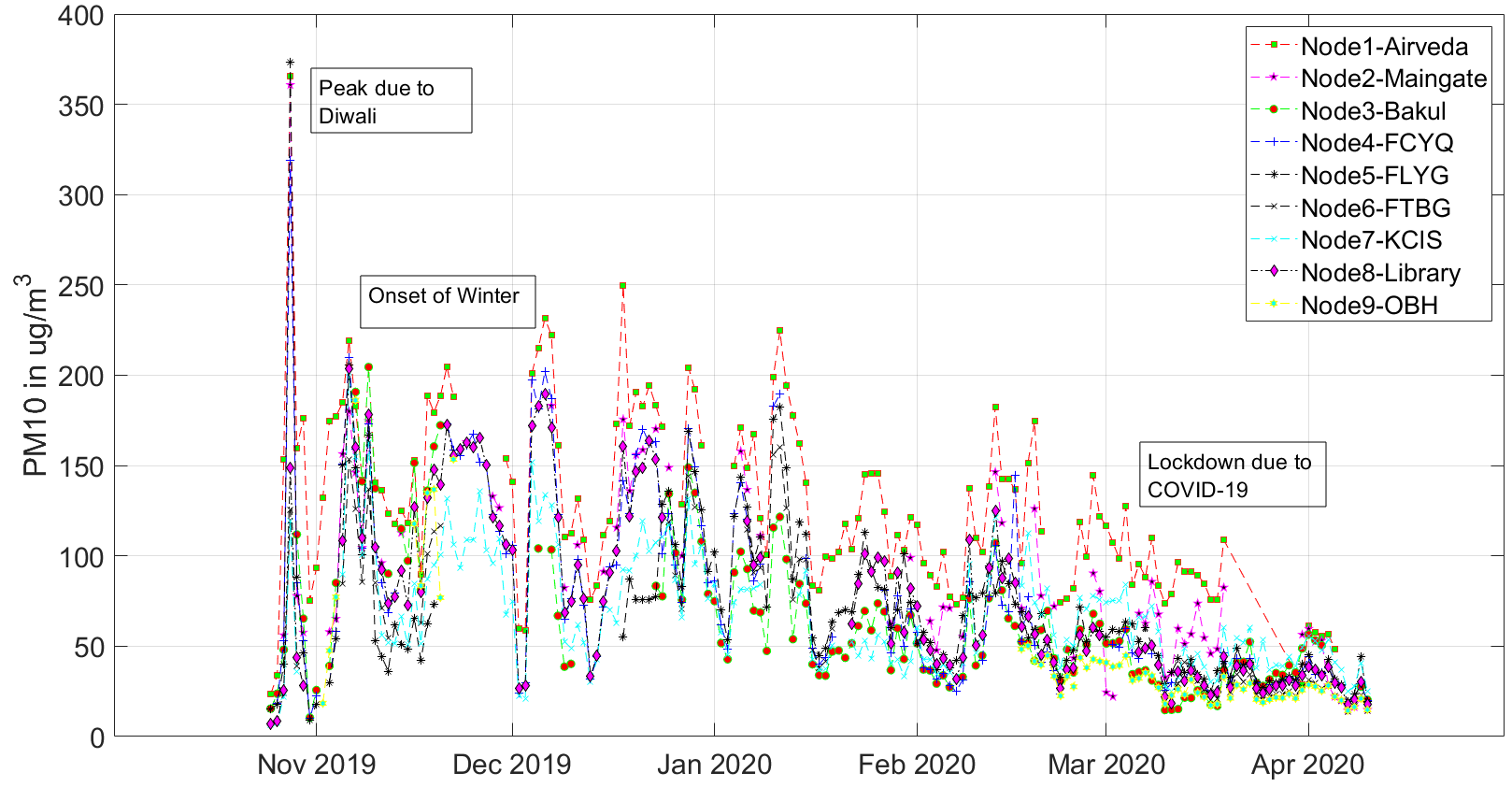}}
    \label{LinePlotPM10}
    }
    \caption{Time Series of PM2.5 and PM10 values (daily averages).}
    \label{Timeseriesplots}
\end{figure*}
Fig. \ref{Timeseriesplots} shows time series plots for nine nodes with daily averaging for both PM2.5 and PM10 samples. The deployment period of the nodes has been more than five months from 26 October 2019 to 10 April 2020. Four important observations can be made from this figure. First, a clear peak can be observed for all nodes on 27 October 2019 resulting from the widespread burning of firecrackers during the celebration of Diwali festival. Dominant peaks can be seen at Node5-FLYG and Node4-FCYQ, where residents of the campus burst crackers. However, the peak in PM values due to fire-crackers died down in next few days. Second observation from the figure is that the PM values again started increasing with the onset of winter in November 2019 and peak in December 2019 and January 2020 during peak winter with temperatures in Hyderabad  between 10-30 \si{\celsius}. Third observation is that as the winter weakens in February 2020, the PM values in general started decreasing. Final observation is that the PM plots show very low values of PM in March and April 2020. This can be attributed to the drastic reduction in traffic and construction activities in and around campus as the State and Central Government gradually started increasing restrictions to prevent spread of Covid-19 from the first week of March and finally declared nationwide lockdown since 22 March 2020 till 30 April 2020. 

\subsection{Correlation Analysis}
\begin{figure}[tbhp]
    \centering
    \subfigure[PM 2.5]
    {\includegraphics[width=0.9\columnwidth]{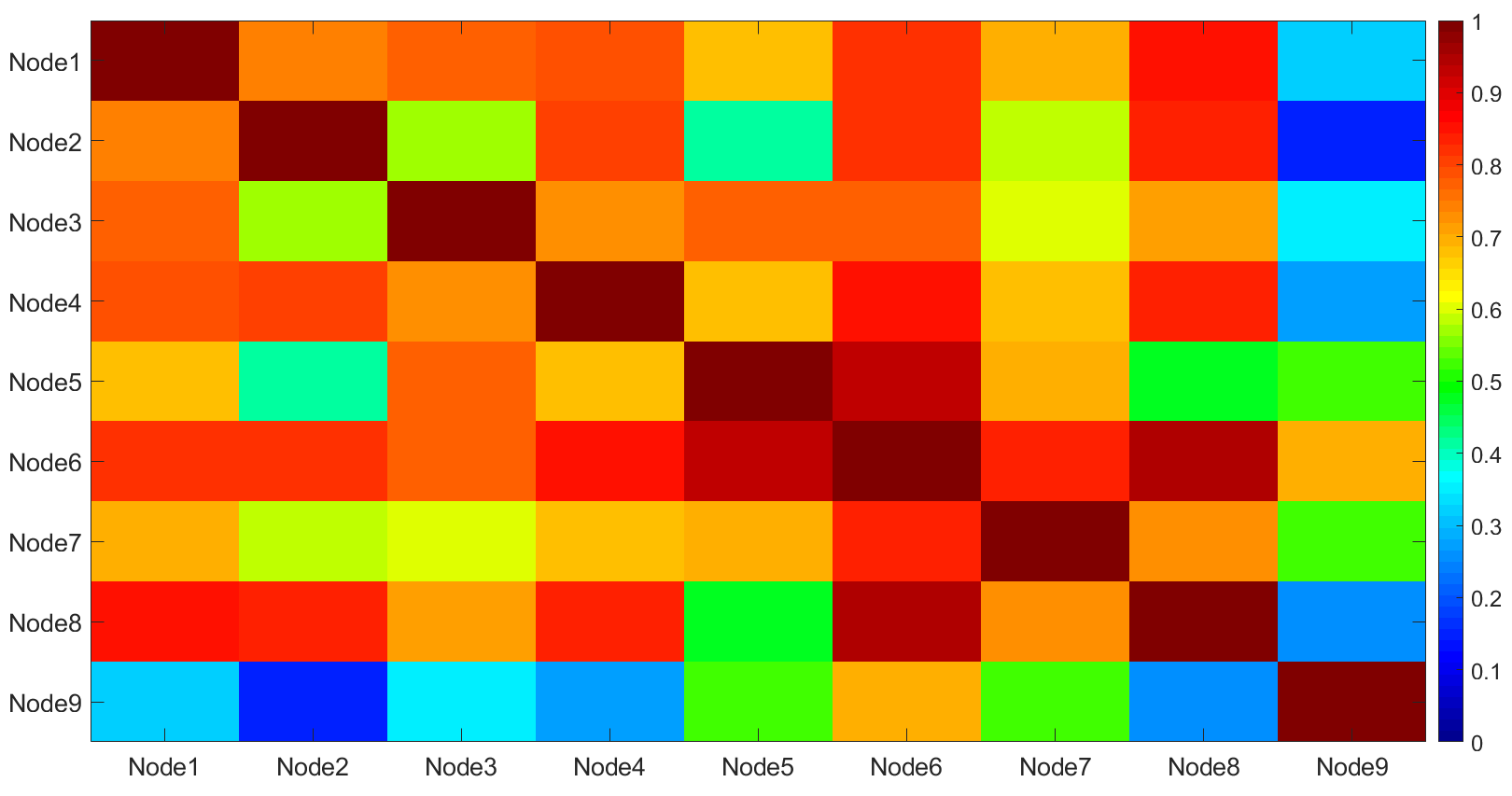}
    \label{K-PM25}
    }
    \\
    \subfigure[PM 10]
    {\includegraphics[width=0.9\columnwidth]{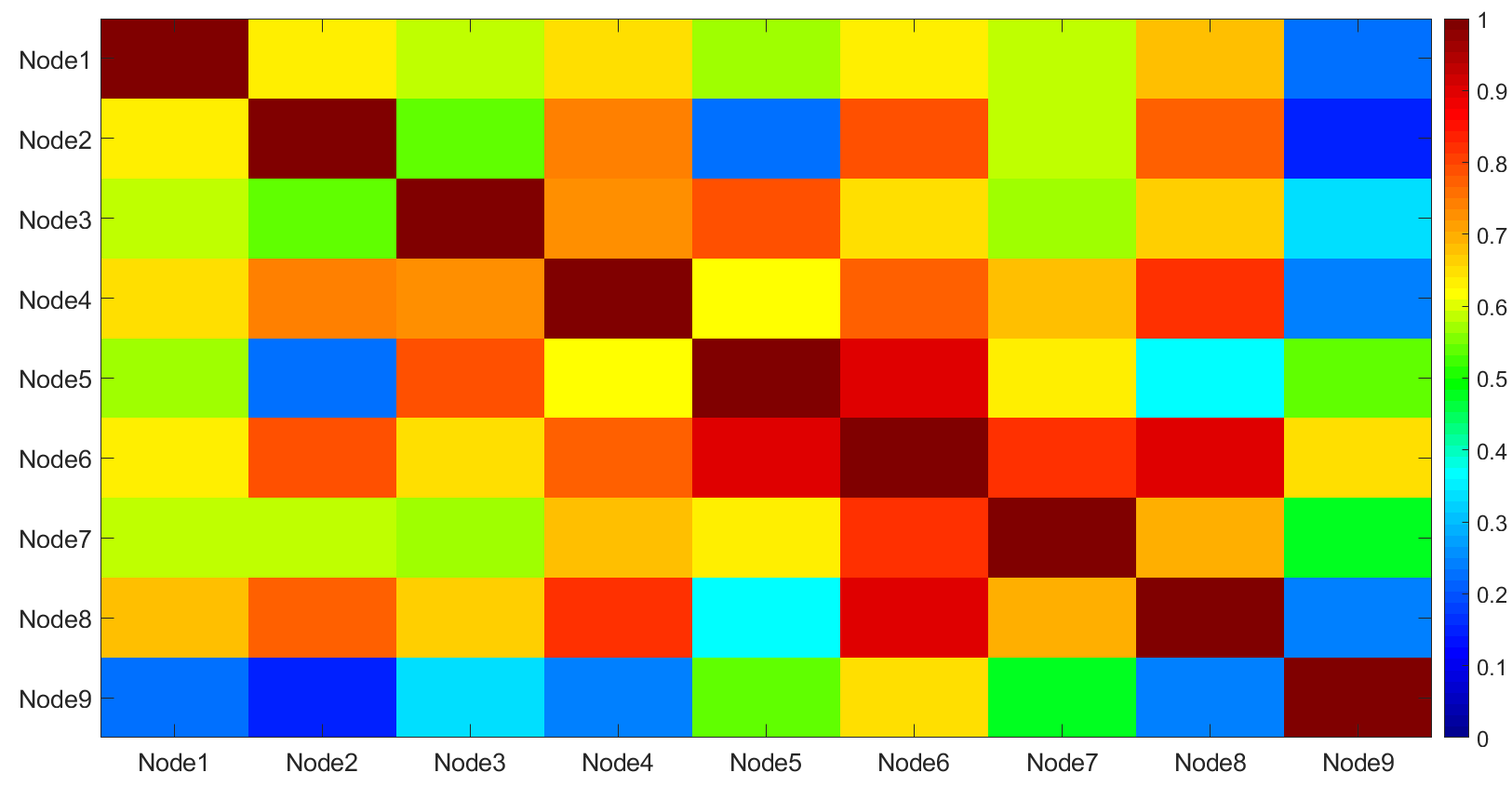}
    \label{K-PM10}
    }
    \caption{Kendall's correlation between the nodes for PM2.5 and PM10.}
    \label{Kendall}
\end{figure}

Kendall's correlation coefficients $\tau$ between the nine sensor nodes have been calculated using five minute averages of PM2.5 and PM10 samples. The values of the Kendall's coefficients are shown in the Fig. \ref{Kendall}. The Kendall's coefficient varies from a value of 0.1538 to 0.9492 for PM2.5 samples and 0.14598 to 0.8954 for PM10 samples. The significant variation between correlation values highlight the spatial variability between the PM values at different nodes. The maximum amount of correlation has been shown by the Node6-FTBG and Node8-Library. The least correlation for both PM2.5 and PM10 is shown by Node2-Maingate and Node9-OBH, which are farthest from each other by about 600 \si{\meter}. Node5-FLYG and Node6-FTBG also show very high amount of correlation of 0.9223 and 0.8926 for PM2.5 and PM10 values, respectively. Node5-FLYG and Node6-FTBG both are placed in  similar geographical conditions (facing an open ground). Node9-OBH shows very less correlation in most of the pairs of nodes, as the node is located far inside the residential block of the campus with close to zero vehicle frequency. 

Kendall's coefficients have also been calculated between the Node1-Airveda and the six CPCB stations deployed across the city and the values are shown in the Table \ref{CPCBcorrelation}. The kendall's coeffiecients vary from as low as 0.17 to maximum being 0.68 in the case of PM2.5 and 0.37 to 0.47 in the case of PM10. In general the values are lesser than 0.5 which implies very weak relationship between the stations. The nearest CPCB station from the Node1-Airveda show Kendall tau values of 0.47 and 0.68 for PM10 and PM2.5 respectively indicating the very low relation between two locations which are approximately only 3 \si{km} apart. This shows the necessity for local PM monitoring for a better understanding of the street-level values.

\begin{table}[bthp]
\caption{Kendall's coefficients of correlation between Node1-Airveda and CPCB stations.}
\centering
\begin{tabular} {|p{1.4cm}|p{0.6cm}|p{0.6cm}|p{0.6cm}|p{0.6cm}|p{0.6cm}|p{0.6cm}|}
 \hline
 \multicolumn{1}{|c|}{Node1-Airveda} & \multicolumn{6}{c|}{Kendall's Coefficients with CPCB stations}                                             \\ [0.5ex] \hline
 PM2.5 & 0.68 & 0.59 & 0.64 & 0.58 & 0.17 &0.50\\ 
 \hline
 PM10 & 0.47 & 0.37 & 0.47 & 0.42 & NA & 0.37\\ [0.5ex]  
 \hline
\end{tabular}
\label{CPCBcorrelation}
\end{table}
\subsection{Spatial Interpolation}
\begin{figure}[htbp]
    \centering
    \subfigure[At 17:00]
    {\includegraphics[width=0.8\columnwidth]{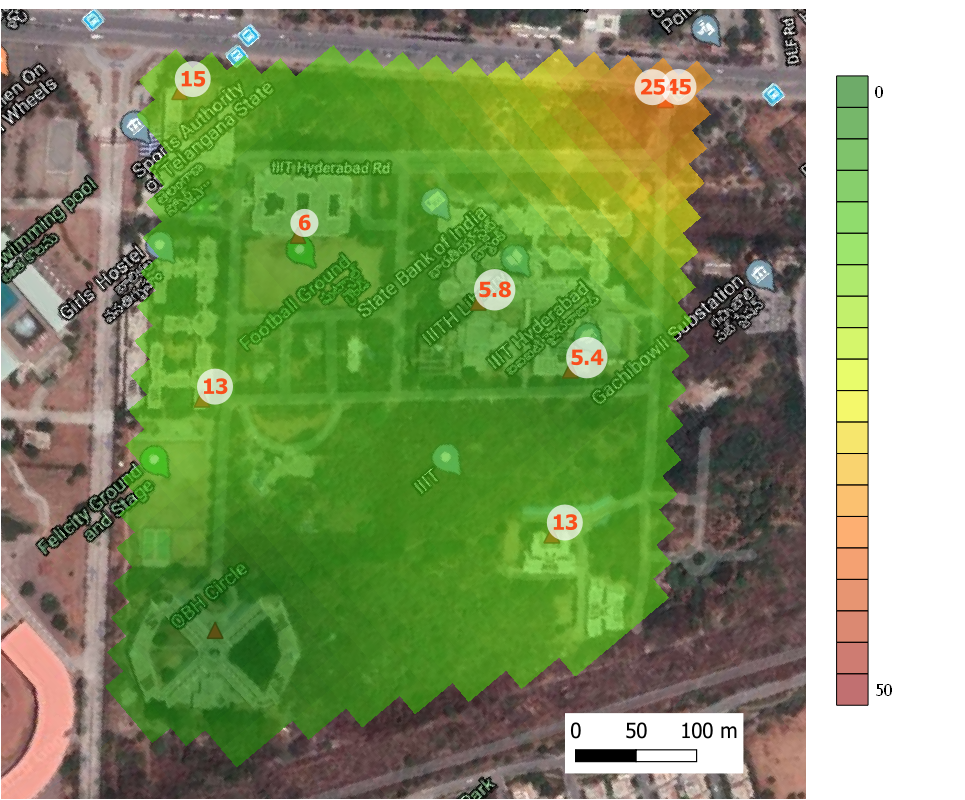}
    \label{IDW1900}
    }
    \quad
    \subfigure[At 22:40]
    {\includegraphics[width=0.8\columnwidth]{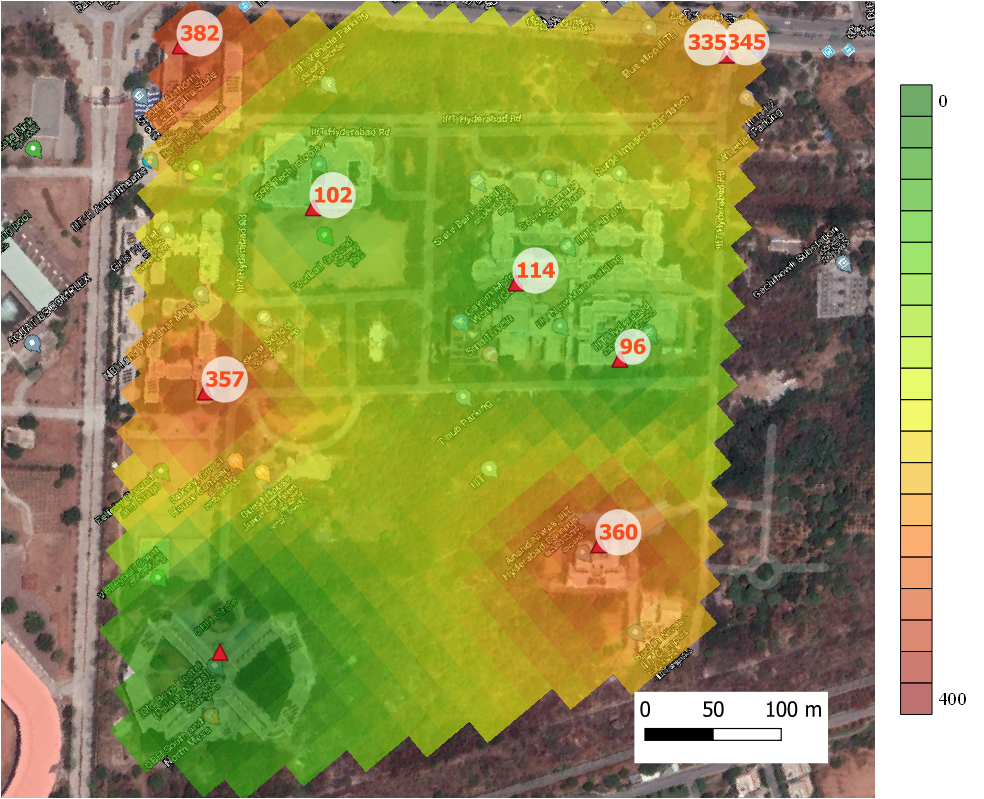}
    \label{IDW2240}
    }
    \caption{Spatial interpolation of PM10 values in IIIT-H campus using IDW at 17:00 hrs and 22:40 hrs on the day of Diwali (27 October 2019). Note the difference in scales in the two maps for convenient viewing.}
    \label{Spatial}
\end{figure}

Figs. \ref{IDW1900} and \ref{IDW2240} show IDW based interpolation maps for PM10 plotted at timestamps 19:00:00 (before burning crackers) and 22:40:00 (after burning crackers) on the day of Diwali. In Fig. \ref{IDW1900}, the hot-spot of the PM10 values is at the Node1-Airveda and Node2-Maingate which are placed near a six-lane highway and exposed directly to vehicular pollution. Spatial variation can be clearly seen in Fig. \ref{IDW1900} between the nine points in an area of only 66 acres (0.267 km2) with Node6-FTBG, Node7-KCIS and Node8-Library showing comparatively lower values being in the center of the campus. In Fig. \ref{IDW2240}, which shows the values at 22:40 after bursting of crackers, the values increase dramatically by 10 to 25 times. Now the number of hot-spots has increased to four, of which Node5-FLYG and Node4-FCYQ are the sites for bursting crackers while Node1-Airveda, Node2-Maingate and Node3-Bakul are affected by both vehicular pollution and crackers burned outside the campus. Node9-OBH was off due to some technical issue on the evening of Diwali, which has affected the interpolated values at that point and resulting in lower values than the actual. Fig. \ref{IDW2240} shows three nodes and the area in the center of the campus which are surrounded by the pollution hot-spots but yet show significantly lower values of PM. The spatial variation within the nine nodes is dominantly seen and hence demonstrates the need for local deployment of sensor nodes for accurate monitoring of the air quality conditions locally. Fig. \ref{Spatial} also show the temporal variation of the values within a small time period of five hours an increase in value from 13 to around 360 at the Node5-FLYG and Node4-FCYQ. Although similar results have been obtained for PM2.5, they are not shown here for brevity.

\section{Conclusion}\label{sec:Conclusion}
In this paper, the dense deployment of IoT nodes has been evaluated for monitoring PM values in urban Indian setting. For this, nine nodes have been deployed in a small campus of IIIT-H. A web-based dashboard has been developed for real time PM monitoring. The measurements done over the period of more than five months clearly show significant increase in PM values during Diwali as well as the noticeable reduction in PM values during national lockdown during COVID-19. It has been shown that correlation coefficient between some nodes in the same campus have low values demonstrating that the PM values across a small region may be significantly different. Moreover, the IDW-based spatial interpolation results on the day of Diwali show significant spatial variation in PM values in the campus ranging from 96 to 382 for locations just a few hundred meters apart for PM10. The results also show notable temporal variations with PM values rising up to 25 times at the same spot in few hours. Thus, there is sufficient motivation to use dense deployment of IoT nodes for improved spatio-temporal monitoring of PM values. 

\bibliographystyle{IEEEbib}
\bibliography{References}

\end{document}